\documentclass[twocolumn]{emulateapj}

\usepackage{amsmath}
\usepackage[title]{appendix}

\shorttitle{White Dwarfs in NGC 2099}
\shortauthors{Cummings et al.}

\begin{document}

\title{Initial-Final Mass Relation for 3 to 4 M$_\odot$ Progenitors\\
 of White Dwarfs from the Single Cluster NGC 2099\textsuperscript{1}}


\author{Jeffrey D. Cummings\altaffilmark{2}, Jason S. Kalirai\altaffilmark{3,2}, P.-E. Tremblay\altaffilmark{3,5}, AND Enrico Ramirez-Ruiz\altaffilmark{4}}
\affil{}

\footnotetext[1]{Based on
observations with the W.M. Keck Observatory, which is operated as a scientific partnership
among the California Institute of Technology, the University of California, and NASA, was made 
possible by the generous financial support of the W.M. Keck Foundation.}

\altaffiltext{2}{Center for Astrophysical Sciences, Johns Hopkins University,
Baltimore, MD 21218, USA; jcummi19@jhu.edu}
\altaffiltext{3}{Space Telescope Science Institute, 3700 San Martin Drive, Baltimore, MD 21218, USA;
jkalirai@stsci.edu, tremblay@stsci.edu}
\altaffiltext{4}{Department of Astronomy and Astrophysics, University of California,
Santa Cruz, CA 95064; enrico@ucolick.org} 
\altaffiltext{5}{Hubble Fellow}

\begin{abstract}
We have expanded the sample of observed white dwarfs in the rich open cluster NGC 2099 (M37) with 
the Keck Low-Resolution Imaging Spectrometer.  Of 20 white dwarf candidates, the 
spectroscopy shows 19 to be true white dwarfs with 14 of these having high S/N.  We find 11 of these 
14 to be consistent with singly evolved cluster members.  They span a mass range of $\sim$0.7 
to 0.95 M$_\odot$, excluding a low-mass outlier, corresponding to progenitor masses of $\sim$3 
to 4 M$_\odot$.  This region of the initial final mass relation (IFMR) has large scatter and a 
slope that remains to be precisely determined.  With this large sample of white dwarfs that 
belong to a single age and metallicity population, we find an initial-final mass relation of 
(0.171$\pm$0.057)M$_{\rm initial}$+0.219$\pm$0.187 M$_\odot$, significantly steeper than the 
linear relation adopted over the full observed white dwarf mass range in many previous studies.  
Comparison of this new relation from the solar metallicity NGC 2099 to 18 white dwarfs in the 
metal-rich Hyades and Praesepe shows that their IFMR also has a consistently steep slope.  This 
strong consistency also suggests that there is no significant metallicity dependence of the IFMR 
at this mass and metallicity range.  As a result, the IFMR can be more reliably determined with 
this broad sample of 29 total white dwarfs giving 
M$_{\rm final}$=(0.163$\pm$0.022)M$_{\rm initial}$+0.238$\pm$0.071 M$_\odot$ from M$_{\rm initial}$
of 3 to 4 M$_\odot$.  A steep IFMR in this mass range indicates that the full IFMR is nonlinear.

\end{abstract}

\section{Introduction}

White dwarfs are by far the most common end point of stellar evolution and are a valuable 
tool for understanding a variety of processes at both the stellar and galactic scales.  A key 
aspect in stellar evolution research is the initial final mass relation (IFMR), where 
white dwarf masses are compared directly to the zero-age main sequence (ZAMS) mass of 
their progenitors.  This provides a direct measurement of the integrated mass loss that occurs 
during a stellar lifetime and its dependence on stellar mass.  The IFMR gives critical 
input for a variety of topics, including stellar feedback in galaxy models (Agertz \& Kravtsov 2014), 
interpreting the white dwarf luminosity function (Catal{\'a}n et~al.\ 2008), providing a way to 
measure the age of the Galactic halo (Kalirai 2013), and predicting Type Ia supernovae rates 
(Pritchet et~al.\ 2008; Greggio 2010).

Weidemann (1977) was the first to analyze the IFMR by comparing mass-loss models (Fusi-Pecci \& 
Renzini 1976) to the observed spectroscopic masses of a few white dwarfs in the nearby Hyades 
and Pleiades clusters.  He concluded that the models greatly underestimated the observed mass 
loss.  Subsequent spectroscopic observations of white dwarfs in nearby open clusters provided 
further constraints on the IFMR (Koester \& Reimers 1981, 1985, 1993, 1996; Reimers \& Koester 
1982, 1989, 1994; Weidemann \& Koester 1983; Weidemann 1987; Jeffries 1997).  These studies 
resulted in a broad but sparsely populated relation (see Weidemann 2000 for review) that shows 
a clear trend with higher mass main-sequence stars producing increasingly more massive white dwarfs.

More recently, the amount of data constraining the IFMR has greatly increased (e.g., Claver et 
al. 2001; Dobbie et~al.\ 2004, 2006; Williams et~al.\ 2004; Kalirai et~al.\ 2005; Liebert et~al.\ 
2005; Williams \& Bolte 2007; Kalirai et~al.\ 2007; Kalirai et~al.\ 2008; Rubin et~al.\ 2008; Kalirai et~al.\ 2009; 
Williams et~al.\ 2009; Dobbie \& Baxter 2010; Dobbie et~al. 2012).  While these newer data still retain the general 
trend observed in the earlier IFMR, the scatter in the relation is significant.  Is this scatter indicative
of stochastic mass loss for stars at the same initial mass, or dependencies on metallicity or environmental
effects, or is this scatter simply the result of systematic differences between these studies? 
One likely contributing systematic is the challenge in defining the ages of the clusters 
these white dwarfs belong to, which creates uncertainty in the calculated lifetimes of their 
progenitor stars.  This is not only due to uncertainty in the isochrone
fitting to photometry itself, in particular for clusters with poorly defined turnoffs and red giant clumps
and branches, but that different studies commonly adopt different model isochrones.
Additionally, as discussed in Salaris et~al.\ (2009), for many studies the input physics for 
the adopted isochrones differ from that of the evolutionary models adopted to calculate the 
white dwarf progenitor masses.  All of these factors systematically affect the adopted initial 
masses of the white dwarfs and can introduce large systematic differences between white dwarfs 
from different clusters and those analyzed by different groups.

Variations in cluster metallicity may also play a role in the scatter associated with the 
IFMR.  The effect of metallicity on the IFMR has been modeled by Marigo \& Girardi (2007) and Meng 
et~al.\ (2008), and they both predict that at the same progenitor masses even moderately metal-poor stars 
([Fe/H]$\sim$\;--0.3) result in more massive white dwarfs (by $\sim$0.05 M$_{\odot}$) in comparison solar 
metallicity.  Observationally, evidence for the IFMR's metallicity dependence has been found in the 
extremely metal-rich cluster NGC 6791 ([Fe/H]=+0.47$\pm$0.04; Gratton et~al.\ 2006), where the 
observed lower-mass white dwarfs appear to be undermassive relative to the field distribution (Kalirai 
et~al.\ 2007).  In contrast to this, we note that for the moderate-mass white dwarfs that we
focus on in our study, both models predict that the IFMR for moderately metal-rich stars ([Fe/H]$\sim$0.15) is 
not meaningfully different from the solar metallicity IFMR, with final masses different by 
less than 0.01 M$_{\odot}$.

First, to limit these effects of varying metallicity and open cluster age uncertainty, we
can analyze a large sample of white dwarfs from a single open cluster that is rich with a well defined
turnoff and red giant clump, giving a detailed isochrone fit.  The intermediate aged (520$\pm$50 Myr; 
Kalirai et al 2001; hereafter K01) open cluster NGC 2099 (M37) is an ideal cluster for this.  
The white dwarfs presented here from NGC 2099 cover masses from 0.7 to 0.95 M$_\odot$, a region of the IFMR that still 
exhibits large scatter and an uncertain slope.  Second, with a concern for robustness, we have compared 
our results to the similar mass range of 18 white dwarfs observed in the Hyades and Praesepe.  
Because these three clusters span a moderate range of metallicity, this also directly tests the metallicity 
dependence of the IFMR in this mass and metallicity range (0.0$\lesssim$[Fe/H]$\lesssim$0.15).  Both the Hyades and
Praesepe provide a valuable reference for our work because their age and metallicites are well determined
and their white dwarfs are bright with high signal to noise (S/N) spectroscopy; we take their parameters from the
consistent analysis performed by Kalirai et~al.\ (2014).  

The structure of this paper is as follows, in Section 2 we discuss both the photometric 
and spectroscopic observations that provide the foundation for this work.  In Section 3 we discuss our
white dwarf spectroscopic analysis.  In Section 4 we discuss our cluster membership determinations.
In Section 5 we discuss the NGC 2099 IFMR and compare to that of the Hyades and Praesepe.  In Section 5 
we discuss our conclusions.  Lastly, the Appendix includes a discussion of the last 20 years of cluster 
parameters analysis for NGC 2099 and justification for our adopted values.

\section{Observations}

Based on the deep photometric observations of NGC 2099 with the Canada-France-Hawaii 
telescope (K01), a sample of 67 white dwarf candidates were found in the central 15' 
in NGC 2099.  From comparison to a nearby off-cluster field, approximately 50 of these
white dwarf candidates were estimated to be cluster white dwarfs.  Observations of
white dwarf candidates have been performed with Keck I using the Low Resolution Imaging 
Spectrometer (LRIS; Oke et~al.\ 1995) and with Gemini using GMOS.  With Keck I, 22 
white dwarf candidate spectra were acquired across two 5'x7' fields using two multi-object
spectrum masks (F1 and F2).  For spectral flux calibration, flux standards were observed each night 
using long-slit spectral observations at the same settings.  During the first observations 
on 2002 December 4, the F1 and F2 masks were each observed for 6000 seconds using the 
600/4000 grism, giving a resolution of $\sim$4 \AA\ and a dispersion of $\sim$0.6 \AA\, 
per pixel, (Program ID: U45L-2002B; PI: B. Hansen, UCLA).\footnote[6]{The 2002 data were 
taken from the public Keck Observatory Archive.}  During the second observations on 2008 December 
23 and 24, the F2 mask only was observed for 9600 seconds the first night and 13200 
seconds the second night.  These 2008 observations used the slightly lower resolution 
400/3400 grism, giving a resolution of $\sim$6 \AA\ and a dispersion of $\sim$1.1 \AA\, 
per pixel, (Program ID: U077LA-2008B; PI: E. Ramirez-Ruiz, UCSC).  Both grisms give spectra
that span a series of five Balmer lines (H$\beta$, H$\delta$, H$\gamma$, H$\epsilon$, and H8)
that are sensitive to stellar temperature and surface gravity.  The observing conditions 
were clear with subarcsecond seeing during 2002 December 4.  The conditions were similarly good 
during 2008 December 23 but relatively poor during 2008 December 24.  

Unlike the F1 mask, which was observed only in 2002, the F2 mask was observed twice with two different
grisms.  The 2008 F2 observations were significantly deeper than the 2002 F2 observations, so we
have focused on the 2008 data and do not attempt to coadd these observations of two different resolutions.  
However, we have used the 2002 F2 data to check for systematic differences in the spectral fits caused
by the two different grisms, and we found that the resulting parameters show no systematic
difference.  Therefore, we can confidently consider our F1 and F2 observations from these two grisms together.

We have reduced and flux calibrated our final F2 and F1 data self-consistently 
using the IDL based XIDL pipeline\footnote[7]{Available at http://www.ucolick.org/$\sim$xavier/IDL/}. 
This consistency further limits potential systematics that could be introduced through using the 
final spectra from Kalirai et~al.\ 2005 (hereafter K05), where the Keck F1 mask observations 
were first presented.  In K05 the original F2 data from 2002 were also presented but were of too low S/N 
for detailed analysis.  For completeness, we note that four white dwarfs from K05 are not presented 
here.  The first being WD11, where its placement on the extreme edge of the F1 mask caused its spectrum 
to lose everything blueward of $\sim$4000 \AA.  Similarly, WD3, WD4, and WD7 were the white dwarfs observed 
only by Gemini/GMOS, which has limited blue sensitivity.  For all four of these white dwarf observations,
H$\epsilon$ and H8 were not available, and because these are the lines most sensitive to 
gravity we have concluded these were not suitable for our current study.  

Across both LRIS masks we have acquired spectra for 20 white dwarf candidates that have sufficient 
S/N to analyze.  One of these candidates, WD8, appears to be consistent with a main 
sequence field dwarf.  The remaining 19 were found to be real white dwarfs, 14 of which have high
enough signal to reliably determine both initial and final masses and cluster membership. 

\pagebreak

\section{Spectroscopic White Dwarf Analysis}

We simultaneously fit the five Balmer lines in each spectrum using the recent models of
Tremblay et~al.\ (2011).  The automated fitting technique is based on that described by Bergeron
et~al.\ (1992), where each line profile is normalized by two defined continua regions surrounding
each line.  This prevents uncertainties in the fit due to flux calibration.  Based on the methods of 
Levenberg-Marquardt (Press et~al.\ 1986), the final fits are 
determined by minimizing the $\chi^2$ when simultaneously considering both T$_{\rm eff}$ and log g. 
Combination in quadrature of the resulting
internal error and the external uncertainties from the data calibration, estimated at 1.2\% in 
T$_{\rm eff}$ and 0.038 dex in log g (Liebert, Bergeron, \& Holberg 2005), are used to estimate our
final uncertainties.  These final results and errors are presented in Table 1\footnote[8]
{Relative to K05, our adoptions of hot vs. cool models around the maximum strength of the Balmers 
lines at T$_{\rm eff}$ = 13,000 K have changed for WD1, WD9, and WD17.}.

Applying each of the final T$_{\rm eff}$ and log g to the cooling models for a carbon/oxygen core 
composition with a thick hydrogen layer by Fontaine et~al.\ (2001) gives the mass, absolute 
V magnitude, intrinsic B-V color, and white dwarf cooling age.  
This composition is appropriate for white dwarfs in this range of T$_{\rm eff}$ and mass, and 
Tremblay \& Bergeron (2008) observationally estimate that $\sim$85\% of DA white dwarfs cooler than
30,000 K have a thick hydrogen layer.  Therefore, these spectroscopic masses provide a direct 
measurement of the
white dwarf's current mass independent of any other assumptions or observational parameters.  

We should consider the S/N (per resolution element) and the resulting 
errors for our white dwarfs (see Table 1).  The WD12, WD14, WD15, WD19, and WD23 spectra have low
S/N and mass errors greater than 0.1 M$_\odot$.  Therefore, their parameters are presented for reference 
but have been cut from our final analysis.  This is because not only are their membership determinations
unreliable, but their large mass errors and more importantly their dramatically increasing cooling age 
uncertainties limit their application to the IFMR.

\begin{center}
\begin{deluxetable*}{l c c c c c c c c}
\multicolumn{9}{c}%
{{\bfseries \tablename\ \thetable{} - Initial and Final Parameters}} \\
\hline
ID&T$_{\rm eff}$&log g&M$_{WD}$   &t$_{cool}$&M$_i$      &M$_{i470}$ &M$_{i570}$&S/N\\
  &(K)      &     &(M$_\odot$)&(Myr)     &(M$_\odot$)&(M$_\odot$)&(M$_\odot$)&\\
\hline
\multicolumn{9}{l}{{NGC 2099 Likely Single Star White Dwarf Members}}  \\
\hline
WD18 & 24900$\pm$600  &  8.21$\pm$0.06 & 0.76$\pm$0.036 &  44$^{+11}_{-10}  $ & 3.00$^{+0.03}_{-0.02}$ & 3.12 & 2.90& 73  \\ 
WD2  & 22200$\pm$650  &  8.24$\pm$0.07 & 0.77$\pm$0.045 &  81$^{+18}_{-16}  $ & 3.09$^{+0.04}_{-0.03}$ & 3.23 & 2.98& 51  \\    
WD24 & 18700$\pm$700  &  8.29$\pm$0.11 & 0.80$\pm$0.068 & 163$^{+40}_{-35}  $ & 3.32$^{+0.15}_{-0.10}$ & 3.52 & 3.17& 39  \\ 
WD6  & 16700$\pm$750  &  8.44$\pm$0.11 & 0.89$\pm$0.069 & 299$^{+73}_{-62}  $ & 3.96$^{+0.66}_{-0.35}$ & 4.37 & 3.67& 29  \\    
WD21 & 16900$\pm$700  &  8.37$\pm$0.11 & 0.85$\pm$0.069 & 258$^{+63}_{-52}  $ & 3.72$^{+0.39}_{-0.23}$ & 4.03 & 3.49& 34  \\     
WD5  & 18100$\pm$650  &  8.21$\pm$0.01 & 0.74$\pm$0.062 & 156$^{+36}_{-32}  $ & 3.30$^{+0.13}_{-0.09}$ & 3.49 & 3.15& 29  \\    
WD10 & 19700$\pm$650  &  8.15$\pm$0.09 & 0.71$\pm$0.054 & 104$^{+25}_{-22}  $ & 3.15$^{+0.07}_{-0.06}$ & 3.29 & 3.03& 32  \\    
WD16 & 17000$\pm$900  &  8.41$\pm$0.15 & 0.87$\pm$0.095 & 269$^{+92}_{-72}  $ & 3.78$^{+0.72}_{-0.33}$ & 4.11 & 3.54& 19  \\    
WD17 & 17700$\pm$1050 &  8.56$\pm$0.16 & 0.96$\pm$0.099 & 311$^{+116}_{-91} $ & 4.05$^{+1.54}_{-0.52}$ & 4.50 & 3.74& 17  \\     
WD13 & 19900$\pm$900  &  8.47$\pm$0.13 & 0.91$\pm$0.082 & 189$^{+57}_{-47}  $ & 3.42$^{+0.24}_{-0.16}$ & 3.62 & 3.25& 24  \\    
WD9  & 16200$\pm$800  &  7.95$\pm$0.14 & 0.59$\pm$0.078 & 139$^{+47}_{-38}  $ & 3.25$^{+0.16}_{-0.11}$ & 3.42 & 3.11& 20  \\
\hline
\multicolumn{9}{l}{{White Dwarfs Inconsistent with Single Star Membership}}  \\
\hline
WD22 & 19900$\pm$550  &  8.48$\pm$0.07 & 0.92$\pm$0.043 & 193$^{+30}_{-27}  $ & --  & -- & -- & 61 \\ 
WD20 & 19400$\pm$650  &  8.24$\pm$0.09 & 0.76$\pm$0.057 & 131$^{+30}_{-26}  $ & --  & -- & -- & 44 \\ 
WD1  & 11100$\pm$250  &  8.22$\pm$0.08 & 0.74$\pm$0.051 & 630$^{+94}_{-84}  $ & --  & -- & -- & 54 \\ 
\hline
\multicolumn{9}{l}{{Low Signal to Noise White Dwarfs}}  \\
\hline
WD23 & 14200$\pm$2300 &  8.36$\pm$0.29 & 0.84$\pm$0.180 & 408$^{+356}_{-203}$ & --  & -- & -- & 13 \\ 
WD15 & 13200$\pm$1000 &  8.50$\pm$0.17 & 0.92$\pm$0.110 & 621$^{+285}_{-192}$ & --  & -- & -- & 15  \\ 
WD14 & 11900$\pm$560  &  8.07$\pm$0.19 & 0.65$\pm$0.110 & 415$^{+146}_{-102}$ & --  & -- & -- & 16  \\ 
WD19 & 20700$\pm$2050 &  8.50$\pm$0.30 & 0.93$\pm$0.180 & 177$^{+143}_{-92} $ & --  & -- & -- & 9.5 \\ 
WD12 & 13400$\pm$1300 &  7.93$\pm$0.23 & 0.57$\pm$0.120 & 250$^{+130}_{-108}$ & --  & -- & -- & 13  \\ 
\hline
\vspace{-0.2cm}  
\end{deluxetable*}
\end{center}

\vspace{-0.9cm}

\section{White Dwarf Cluster Membership}

To learn the formation history of these white dwarfs, we must first consider their 
membership of NGC 2099.  The membership is based on our spectroscopic white dwarf parameters 
in comparison to the observed photometry and the adopted cluster parameters.  For this study 
we adopt the photometry and cluster parameters from K01, which are E(B-V)=0.21$\pm$0.03, 
(m-M)$_V$=11.55$\pm$0.13, an age of 520 Myr, and a solar metallicity.  This cluster does 
have a rich history of photometric analysis, and in the Appendix we discuss the multiple 
studies covering NGC 2099 from the past 20 years and the justification of our adopted 
parameters.  However, in summary, the reddening and age have the most significant variation 
across the 11 studies (see Table 3 in the Appendix).  It is challenging to explain the full 
variation in determined reddenings (E(B-V) from 0.21 to 0.3), but the effects of differing 
adopted metallicity, differing isochrones, and that most studies did not account for the 
B-V dependence of reddening (see Fernie 1963 and Twarog et~al.\ 1997) can explain much of it.  
The large variation in ages (320 to 650 Myr) can be explained by the systematic 
differences between the adopted isochrones in each of the studies, with a minor dependence 
on the difference in adopted reddening and metallicity.  In Figure 1 we illustrate three 
of these important isochrones by fitting the K01 photometry and applying their cluster 
reddening, distance modulus, and metallicity with the applied color dependence of reddening 
from Fernie (1963).  The left panel shows our 520 Myr fit using isochrones from Ventura 
et~al.\ (1998), the middle panel shows a 540 Myr fit using the PARSEC isochrones from 
Bressan et~al.\ (2012) using version 1.2S, and the right panel shows a 445 Myr fit using 
isochrones from Girardi et~al.\ (2000).  This shows that the Girardi et~al.\ (2000) isochrones 
and the very similar Bertelli et~al.\ (1994) isochrones find systematically younger cluster 
ages, and the $\sim$0.06 increase in the B-V color separation of the turnoff and red clump 
illustrates why many studies must adopt a higher reddening to fit the observed turnoff
with these two isochrones.  

\begin{figure*}[!ht]
\begin{center}
\includegraphics[clip, scale=0.9]{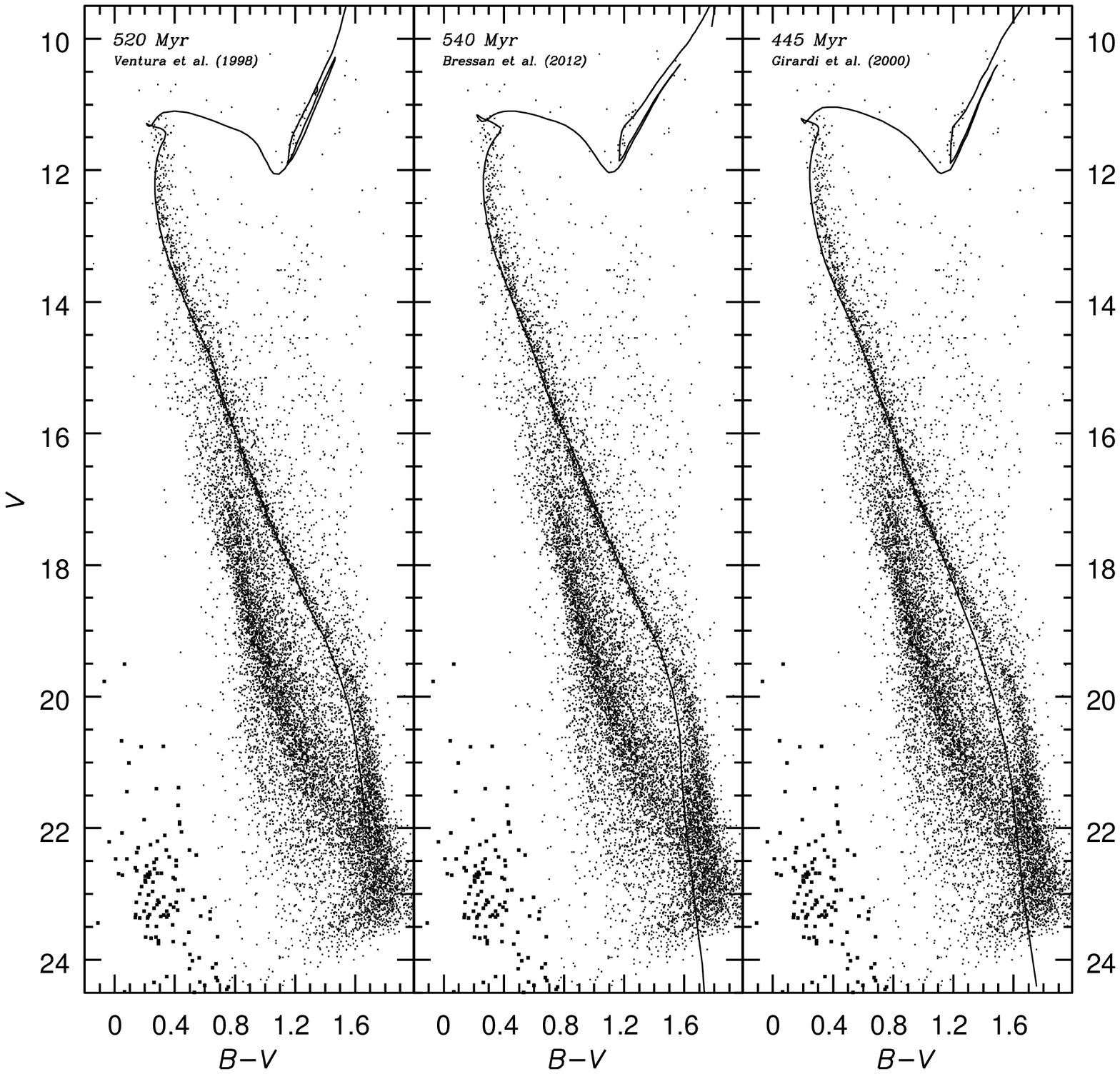}
\end{center}
\vspace{-0.3cm}
\caption{We show three panels of isochrone fits of the K01 photometry for NGC 2099, where we have
also adopted their photometric parameters but apply a color dependence of the reddening.  
(The rich white dwarf population is emphasized with bold data points.)  The 
left and middle panels show our age fits using the quite similar Ventura et~al.\ (1998) and
Bressan et~al.\ (2012) isochrones.  The right panel helps to illustrate the important systematics
introduced by adopting the Girardi et~al.\ (2000) isochrones, which are also nearly equivalent
to the commonly used Bertelli et~al.\ (1994) isochrones.}
\end{figure*}

Table 2 presents the coordinates of our observed white dwarf candidates and the parameters
used for analyzing cluster membership.  We first 
directly compare the observed V magnitude for these white dwarfs to their theoretical absolute magnitude.  
With such a large white dwarf sample from a single cluster, we can comment on the apparent distance 
modulus.  In the left panel of Figure 2, we plot the distribution with 1$\sigma$ error bars of apparent 
distance modulus for each white dwarf.  This is independent of our adopted cluster parameters, and we 
find a majority of the white dwarfs are consistent.  This is based on 
considering the combination in quadrature of the observed and theoretical magnitude errors (1$\sigma$), 
and we define outliers as those that are more than 2$\sigma$ from the final sample mean, which is 
(m-M)$_V$=11.69$\pm$0.18 (solid vertical line).  In Figure 2 we have ordered the white dwarfs accordingly where 
the top 11 are those with consistent distance moduli, the following 3 (WD1, WD20, and WD21) 
are inconsistent and suggest nonmembership, and the final 5 are the low S/N white dwarfs 
shown for comparison.  This white dwarf based distance modulus is consistent with our adopted value of 
(m-M)$_V$=11.55$\pm$0.13, and based on the color dependence of reddening and hence extinction, this 
agreement improves because a moderately larger value (11.58$\pm$0.13; the dashed vertical line) is expected 
for the very blue white dwarfs (see discussion of the color dependence of reddening and 
extinction in the Appendix).  The errors of these distance moduli are illustrated at the top of Figure 2.

\begin{figure*}[!ht]
\begin{center}
\includegraphics[clip, scale=1]{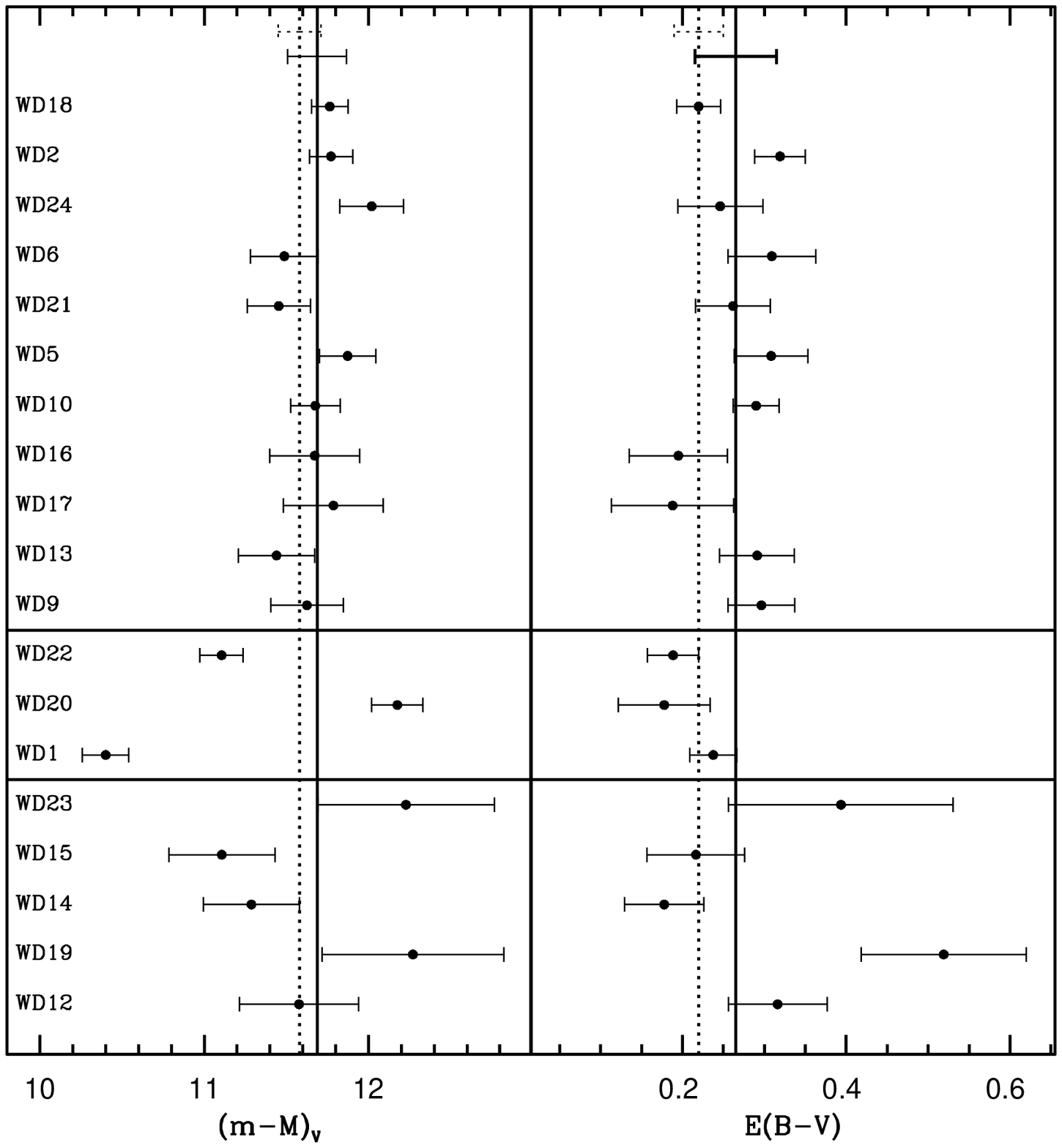}
\end{center}
\vspace{-0.3cm}
\caption{In the left panel we display the distribution and 1$\sigma$ error bars for the white
dwarf based distance moduli.  The upper 11 white dwarfs are consistent (within 2$\sigma$) with
the mean, and we find the middle 3 are inconsistent.  For reference the bottom 5 white dwarfs 
are those that have been cut for their low S/N.  The solid vertical line represents the mean 
of the member distance
moduli and the dashed line is our photometric based distance modulus, with their respective
error bars at the top.  In the right panel we display the distribution and error bars
for the white dwarf based reddening.  Assuringly, the 11 white dwarfs with consistent distance
moduli also have consistent reddenings.  Again, the solid vertical line represents the mean of
the member reddenings, and the dashed line is our photometric based reddening.}
\end{figure*}

In the right panel of Figure 2, we similarly plot the distribution of apparent
reddenings for each white dwarf based on the difference between the theoretical and observed color.
We test for color outliers by considering again the combination in quadrature of the observed and 
theoretical color errors (1$\sigma$), and outliers are those that are more than 2$\sigma$ from 
the final sample mean, which is E(B-V)=0.265$\pm$0.05 (solid vertical line).  Remarkably consistent 
to the magnitude comparison, we find that all white dwarfs with consistent distance moduli
also have consistent reddenings, while one of the three white dwarfs with inconsistent distance moduli
also has inconsistent reddening.  Due to the relatively large photometric color errors, this test is 
not as stringent as the magnitude test, but it is still important.  This white dwarf based reddening 
is in agreement with our adopted reddening of 0.21$\pm$0.03, and the agreement improves further 
because a moderately larger apparent reddening (0.22$\pm$0.03; the dashed vertical line) is expected 
for white dwarfs.

While the numbers are limited, we can also separate the final white dwarfs into their respective
F1 and F2 fields to test for potentially large variations in reddening and extinction, which should be 
considered in clusters of moderate reddening.  The difference in each field's mean reddening
or distance modulus is not found to be significant ($\Delta$E(B-V)=0.02$\pm$0.06; $\Delta$(m-M)$_V$=0.05$\pm$0.25), 
and any potential variable reddening is likely not significant within the relatively small FOV of LRIS itself 
(5'x7').  Therefore, variable reddening or extinction likely have not affected our membership analysis.
We should also note that if we adopt these white dwarf based distance modulus and reddening for our photometric 
isochrone fits, there is only a moderate effect on the age giving a best fit of 490 Myr adopting the Ventura et~al.\ 
(1998) isochrones and 510 Myr adopting the Bressan et~al.\ (2012) isochrones ($\Delta$Age$\sim$30 Myr).

Figure 3 shows the Balmer-line fits to the spectra for these 11 member white dwarfs used in our 
analysis of NGC 2099.  Figure 4 focuses on the 
photometric characteristics of the observed white dwarfs.  Here we have indicated membership, including 
observed photometric error bars for the final members, and have plotted them relative to a series of 
white dwarf cooling models of different mass.

\begin{figure*}[!ht]
\begin{center}
\includegraphics[clip, scale=0.9]{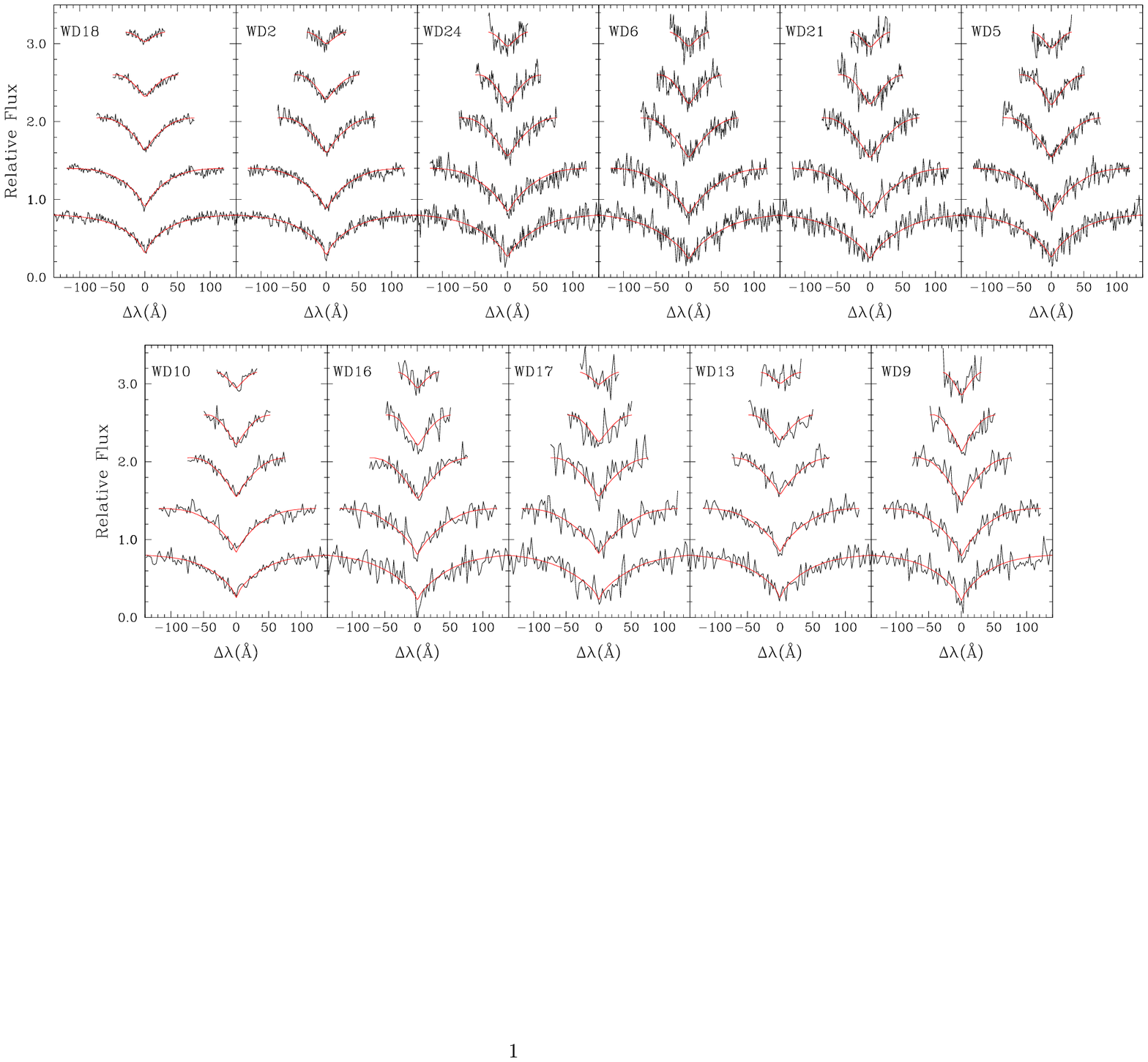}
\end{center}
\vspace{-0.3cm}
\caption{The Balmer fits for the 11 final white dwarf members of NGC 2099.  The H$\beta$, H$\delta$, 
H$\gamma$, H$\epsilon$, and H8 fits are shown from bottom to top.  The upper 6 white dwarf spectra are
from the F2 mask and the bottom 5 white dwarf spectra, which have been binned for display, are from 
the F1 mask.}
\end{figure*}

\begin{figure}[!ht]
\begin{center}
\includegraphics[scale=0.45]{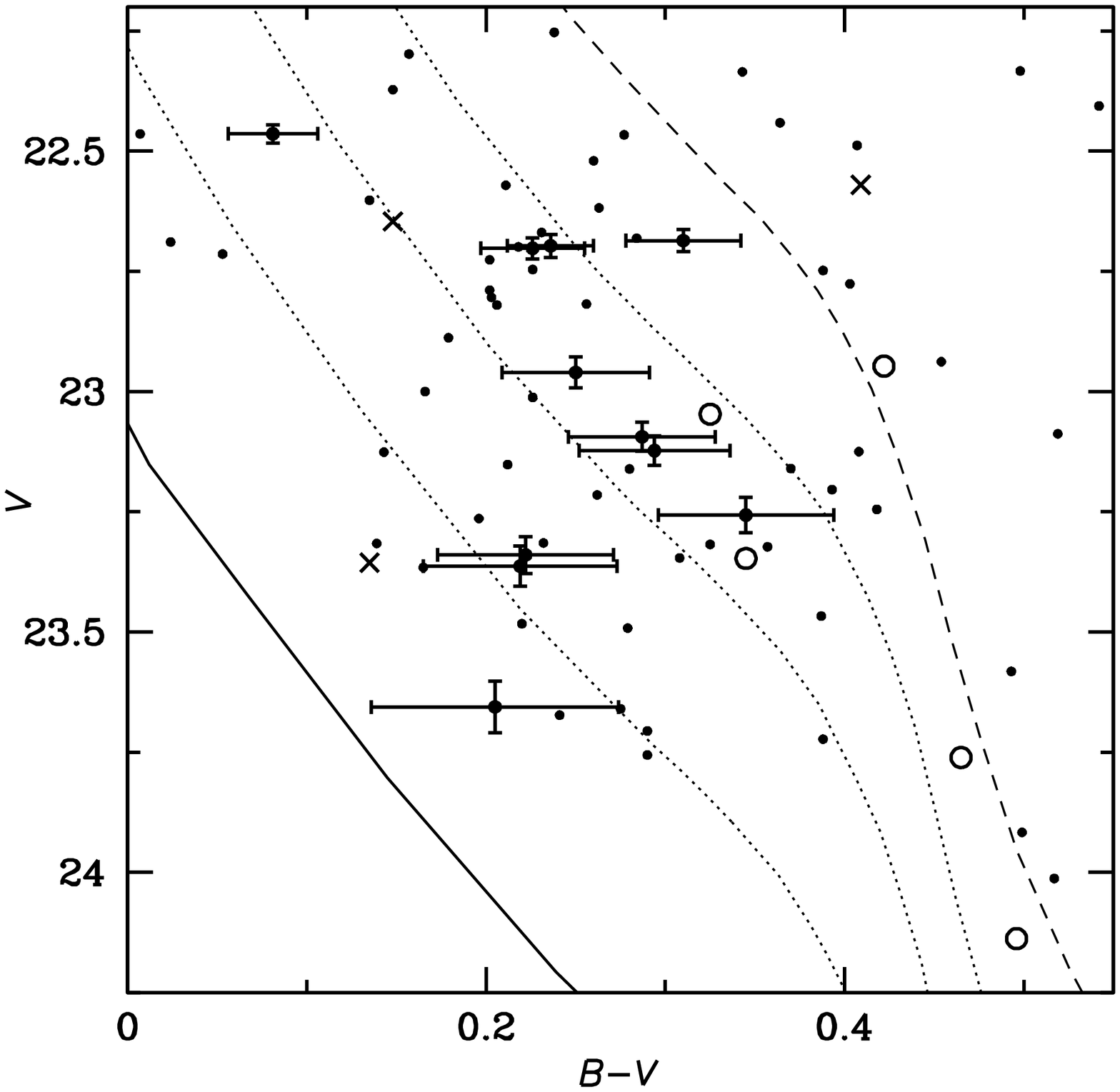}
\end{center}
\vspace{-0.3cm}
\caption{The white dwarf photometry where we differentiate between observed non-members 
(x's), likely singly evolved members (solid circle), observed white dwarfs with low S/N (open circles), 
and unobserved candidates (small points).  White dwarf cooling models are 
plotted for reference from left to right of 1.2 M$_\odot$ (solid line), 1.0, 0.8, and 0.6 
M$_\odot$ (dotted lines), and 0.4 M$_\odot$ (dashed line).  The mean distance modulus and reddening
derived from these white dwarf members are used on these cooling models.}
\end{figure}

In K01 their photometric comparison of white dwarf candidate density within and outside 
the cluster field found that $\sim$75\% of the white dwarf candidates are 
cluster members.  We find that 11 (73\%) of our 14 observed white dwarfs candidates that have S/N
appropriate for membership analysis are consistent with singly evolved NGC 2099 members.
This reconfirms the richness of the WD population of NGC 2099 and the high return on its candidate
observations, where many of the candidates found in the K01 photometry still remain unobserved.

We have determined cluster membership, and this provides a valuable reference for further analysis
of these white dwarfs.  The white dwarf cooling age represents the time since that white dwarf formed 
at the tip of the asymptotic giant branch (AGB).  Therefore, direct comparison of a white dwarf's 
cooling age to its parent cluster's total age gives us the total evolutionary lifetime (to the tip of 
the AGB) for that white dwarf's progenitor star.  We have applied these lifetimes to the models of 
Hurley et~al.\ (2000) at solar metallicity
to determine the ZAMS masses for each progenitor, which can be compared directly to their final observed
white dwarf mass.  Before continuing, Salaris
et~al.\ (2009) noted that inconsistencies between the input physics of isochrones to determine the cluster
age and the input physics of the adopted model for determining evolutionary times at specific progenitor
masses are not always consistent.  However, we find that the progenitor masses found for the stars
at the tip of the AGB at a given age are consistent to within 1\% for Hurley et~al.\ (2000) and both
the Ventura et~al.\ (1998) and Bressan et~al.\ (2012) isochrones at our progenitor mass range.  

\tablefontsize{\footnotesize}
\begin{center}
\begin{deluxetable*}{l c c c c c c c c}
\multicolumn{9}{c}%
{{\bfseries \tablename\ \thetable{} - Membership Test}} \\
\hline
ID&Field&$\alpha$&$\delta$&V       &V     &B-V   &B-V   &t$_{cool}$\\
  &     &(J2000) &(J2000) &(Theory)&(Obs.)&(Theory)&(Obs.)&(Myr)\\
\hline
\multicolumn{9}{l}{{NGC 2099 Likely Single Star White Dwarf Members}}  \\
\hline
WD18  & F2 & 5:52:13.32 & 32:28:55.9& 22.39$\pm$0.11 & 22.464$\pm$0.019 & 0.126$\pm$0.009 & 0.081$\pm$0.025 &  44$^{+11}_{-10} $  \\  
WD2   & F2 & 5:52:17.54 & 32:29:03.4& 22.62$\pm$0.13 & 22.702$\pm$0.022 & 0.172$\pm$0.011 & 0.226$\pm$0.029 &  81$^{+18}_{-16} $  \\ 
WD24  & F2 & 5:52:25.34 & 32:29:35.2& 23.01$\pm$0.19 & 23.340$\pm$0.038 & 0.241$\pm$0.017 & 0.222$\pm$0.049 & 163$^{+40}_{-35} $  \\ 
WD6   & F2 & 5:52:18.44 & 32:29:37.6& 23.46$\pm$0.20 & 23.257$\pm$0.037 & 0.301$\pm$0.022 & 0.345$\pm$0.049 & 299$^{+73}_{-62} $  \\ 
WD21  & F2 & 5:52:26.42 & 32:30:04.3& 23.33$\pm$0.19 & 23.094$\pm$0.030 & 0.290$\pm$0.020 & 0.287$\pm$0.041 & 258$^{+63}_{-52} $  \\ 
WD5   & F2 & 5:52:25.46 & 32:30:54.4& 22.94$\pm$0.17 & 23.123$\pm$0.031 & 0.251$\pm$0.016 & 0.294$\pm$0.042 & 156$^{+36}_{-32} $  \\ 
WD10  & F1 & 5:52:25.82 & 32:36:03.8& 22.71$\pm$0.15 & 22.697$\pm$0.024 & 0.211$\pm$0.014 & 0.236$\pm$0.024 & 104$^{+25}_{-22} $  \\ 
WD16  & F1 & 5:52:22.89 & 32:36:29.4& 23.38$\pm$0.27 & 23.363$\pm$0.042 & 0.289$\pm$0.026 & 0.219$\pm$0.054 & 269$^{+92}_{-72} $  \\ 
WD17  & F1 & 5:52:29.29 & 32:37:06.7& 23.56$\pm$0.30 & 23.656$\pm$0.054 & 0.282$\pm$0.029 & 0.205$\pm$0.069 & 311$^{+116}_{-91}$  \\ 
WD13  & F1 & 5:52:16.21 & 32:38:18.5& 23.21$\pm$0.23 & 22.960$\pm$0.032 & 0.224$\pm$0.020 & 0.250$\pm$0.041 & 189$^{+57}_{-47} $  \\ 
WD9   & F1 & 5:52:25.14 & 32:40:03.6& 22.75$\pm$0.22 & 22.686$\pm$0.023 & 0.279$\pm$0.025 & 0.310$\pm$0.032 & 139$^{+47}_{-38} $  \\ 
\hline
\multicolumn{9}{l}{{White Dwarfs Inconsistent with Single Star Membership}}  \\
\hline
WD22  & F2 & 5:52:04.99 & 32:25:40.6& 23.23$\pm$0.13 & 22.646$\pm$0.023 & 0.224$\pm$0.012 & 0.148$\pm$0.029 & 193$^{+30}_{-27}  $ \\ 
WD20  & F2 & 5:52:12.88 & 32:29:50.0& 22.87$\pm$0.15 & 23.356$\pm$0.044 & 0.222$\pm$0.015 & 0.135$\pm$0.054 & 131$^{+30}_{-26}  $ \\ 
WD1   & F1 & 5:52:26.80 & 32:30:46.7& 23.86$\pm$0.14 & 22.570$\pm$0.020 & 0.437$\pm$0.006 & 0.409$\pm$0.028 & 630$^{+94}_{-84} $  \\ 
\hline
\multicolumn{9}{l}{{Low Signal White Dwarfs and a Non White Dwarf}}  \\
\hline
WD23  & F2 & 5:52:03.30 & 32:27:30.4& 23.60$\pm$0.53 & 24.138$\pm$0.090 & 0.367$\pm$0.061 & 0.496$\pm$0.123 & 408$^{+356}_{-203}$ \\ 
WD8   & F2 & 5:52:15.87 & 32:27:53.4&        --      & 21.655$\pm$0.011 &        --       & 0.421$\pm$0.016 &         --          \\ 
WD15  & F1 & 5:52:27.42 & 32:35:50.0& 23.93$\pm$0.32 & 23.347$\pm$0.042 & 0.394$\pm$0.021 & 0.345$\pm$0.056 & 621$^{+285}_{-192}$ \\ 
WD14  & F1 & 5:52:06.43 & 32:38:29.4& 23.45$\pm$0.29 & 23.047$\pm$0.034 & 0.412$\pm$0.015 & 0.325$\pm$0.046 & 415$^{+146}_{-102}$ \\
WD19  & F1 & 5:52:01.12 & 32:39:50.4& 23.18$\pm$0.55 & 23.761$\pm$0.067 & 0.211$\pm$0.043 & 0.465$\pm$0.091 & 177$^{+143}_{-92} $ \\ 
WD12  & F1 & 5:52:11.96 & 32:40:39.5& 23.06$\pm$0.36 & 22.947$\pm$0.032 & 0.371$\pm$0.041 & 0.422$\pm$0.044 & 250$^{+130}_{-108}$ \\
\hline
\vspace{-0.2cm}
\end{deluxetable*}
\end{center}

\vspace{-1cm}
\section{Initial Final Mass Relation}

Our 11 white dwarfs consistent with singly evolved cluster membership span a broad range of masses 
from 0.59 to 0.96 M$_\odot$.  Figure 5 presents the IFMR for the cluster when adopting 
the age of 520 Myr.  For reference, the linear fit IFMR observed by Kalirai et~al.\ (2009; hereafter 
K09) is plotted with its errors (solid and dashed lines, respectively).  This K09 relation is 
\vspace{-0.1cm}
\begin{equation*} 
M_{\rm final}=(0.101\pm0.006)M_{\rm initial}+0.463\pm0.018 M_\odot,
\vspace{-0.15cm}
\end{equation*}
which is an update to that of Kalirai et~al.\ (2008) that has
additional low-mass data from M4 and updated spectral models more consistent with those adopted here. 
Additionally, these K09 data include the previous NGC 2099 results from K05 and are based on a 
large data set spanning a progenitor mass range of $\sim$ 1 to 6 M$_\odot$.  In Figure 5 a vertical solid
line is shown that represents the ZAMS mass (2.91 M$_\odot$) of a star that is currently on the 
tip of the AGB at 520 Myr.  Therefore, under the adopted age and mass-loss rates we cannot reach 
progenitor masses lower than this in NGC 2099.  For our data the white dwarf mass errors are those 
already described, while the progenitor mass errors are based on the error in white dwarf cooling 
ages.  Because these white dwarfs come from the same cluster, the error in cluster age is systematic 
and not shown in these error bars.  These progenitor parameters are also presented in Table 1
for direct comparison to their final white dwarf parameters.

Based on the somewhat meaningful variation in white dwarf mass errors (nearly a factor of 3), we present
both an unweighted and weighted linear fit to the IFMR for the M$_{\rm initial}$ range of 3 to
4 M$_\odot$.  Figure 5 presents the weighted linear fit (dotted line) given here
\vspace{-0.1cm}
\begin{equation*} 
M_{\rm final}=(0.171\pm0.057)M_{\rm initial}+0.219\pm0.187 M_\odot.
\vspace{-0.15cm}
\end{equation*}
For comparison, the unweighted linear fit of these IFMR data is 
M$_{\rm final}$=(0.218$\pm$0.067)M$_{\rm initial}$+0.051$\pm$0.234 M$_\odot$.  Before we compare the
coefficients, we should clarify that the zeropoint error is the result of propagating the 
slope error to the y-axis, and so they should not be considered independently.  Even under this
limitation, the weighted and unweighted coefficients are consistent within the errors, but we 
take the weighted fit as our final result because we feel the discrepant WD9 and WD13 overly 
influence the slope of the unweighted relation.

WD9 and WD13 are the only two members that are greater than 1$\sigma$ (white 
dwarf mass error) from the
weighted linear fit.  WD13 is 1.35 $\sigma$ discrepant and WD9 is 2.42 $\sigma$ discrepant from 
the relation.  WD13's minor discrepancy is not of significant
concern, and its high mass (0.91 M$_\odot$) makes it unlikely that it is a field dwarf contaminating 
our sample.  However, WD9 is far more discrepant, and its low mass (0.59 M$_\odot$) is more 
difficult to explain, where when assuming standard stellar evolution a progenitor star that 
would create a white dwarf of this mass would take several Gyr to fully evolve.  The effects of
a binary companion could possibly explain the significant mass loss that would be required to create this
white dwarf, but no evidence for a binary companion was found in its magnitude and color analysis.
It is more probable that this white dwarf is simply contaminating our cluster membership sample rather 
than it being indicative of any significant variation of the mass-loss history of a singly evolved
star at progenitor mass of 3.25 M$_\odot$.  Its low white 
dwarf mass is also comparable with field white dwarf expectations, where the mean mass of white dwarfs 
in the SDSS DR7 sample is 0.598 M$_\odot$ (at T$_{\rm eff}$$>$13,000 K; Kleinman et~al.\ 2013).  We
do not remove WD9 or WD13 from our final sample, but due to their higher errors, their effect on the 
final weighted relation is minimal.

Besides these two white dwarfs, this weighted linear relationship describes the remaining
9 member white dwarfs very well and the errors can fully explain the observed spread in the IFMR. 
Therefore, within this single cluster, there is no convincing evidence for stochastic mass loss
that could explain the large scatter in the IFMR typically observed when combining data from
multiple studies.

Lastly, in Figure 5 the effects of cluster age errors on progenitor mass are 
illustrated.  The younger cluster age (470 Myr) is represented by inverted triangles
and the older cluster age (570 Myr) is represented by triangles.  This shows that in the lower-mass 
white dwarfs (0.7 to 0.8 M$_\odot$) the progenitor masses are only moderately affected by adopting
a 50 Myr older or younger cluster age.  However, the higher-mass stars with more rapid evolutionary
times are increasingly more sensitive to the adopted cluster age.  Therefore, the younger cluster 
age finds a slightly shallower IFMR, but the older cluster age further increases its slope.

Our final weighted IFMR for our analyzed mass range (M$_{\rm initial}$ of $\sim$3 to 4 M$_\odot$) is 
steeper than the linear relation from K09 based on a much broader sample (M$_{\rm initial}$ 
of $\sim$1 to 6 M$_\odot$).  This suggests that a linear relation is insufficient to describe this full mass 
range.  It is also important to clarify, however, that we are comparing to the final linear IFMR fit in 
K09, but their white dwarf data itself do indicate possible nonlinearity across their full mass range.  
In particular, for our mass range analyzed in this current study the K09 data suggested a possibly steeper IFMR slope 
in this region followed by a shallower slope at higher masses, but as discussed in Kalirai et~al.\ (2008) their errors 
and numbers were too limited for a convincing nonlinear fit.  Similar to K09, other recent 
IFMR studies have also considered the possible nonlinearity in the IFMR, but a majority of these have stated that
their errors, 
which resulted primarily from systematics between the differing studies and clusters considered, were 
too significant to justify a nonlinear fit.  For comparison, Ferrario et~al.\ (2005) linearly fit their
data spanning M$_{\rm initial}$ of 2.5 to 6.5 M$_\odot$ with a slope of 0.10038$\pm$0.00518 and a 
y intercept of 0.43443$\pm$0.01467, very similar to the relation from K09.  Salaris et~al.\ 
(2009) linearly fit their data approximately spanning M$_{\rm initial}$ of 1.7 to 7 M$_\odot$
with a slope of 0.084 and a y intercept of 0.466.  Williams et~al.\ (2009) span the same mass
range and find a linear fit with slope of 0.129$\pm$0.004 and y intercept of 0.339$\pm$0.015.
All of these fits are comparable to K09.

Previous evidence in support of a turnover in the IFMR at higher progenitor masses is 
primarily theoretical.  In both IFMR models by Marigo \& Girardi (2007) and Meng et~al.\ (2008) 
a turnover in the slope of the IFMR is predicted near a progenitor mass of $\sim$4 M$_\odot$.  Meng et~al.\ 
(2008) describe this turnover as a result of only these higher-mass stars undergoing a second dredge-up 
event during 
their AGB evolution, which reduces their core mass.  Previous observational evidence for this
turnover is found in Dobbie et~al.\ (2012), where they observed seven white dwarf members of NGC 
3532 spanning progenitor masses of $\sim$3.5 to 6 M$_\odot$.  Similar to our current work, their 
focus on a single rich cluster prevented systematics that could be the source of any of
features in their IFMR.  While their numbers were still too limited to be able to define
two independent slopes both above and below this $\sim$4 M$_\odot$ turning point, the data do suggest 
a turnover exists near this M$_{\rm initial}$.

\begin{figure*}[ht]
\begin{center}
\includegraphics[scale=0.94]{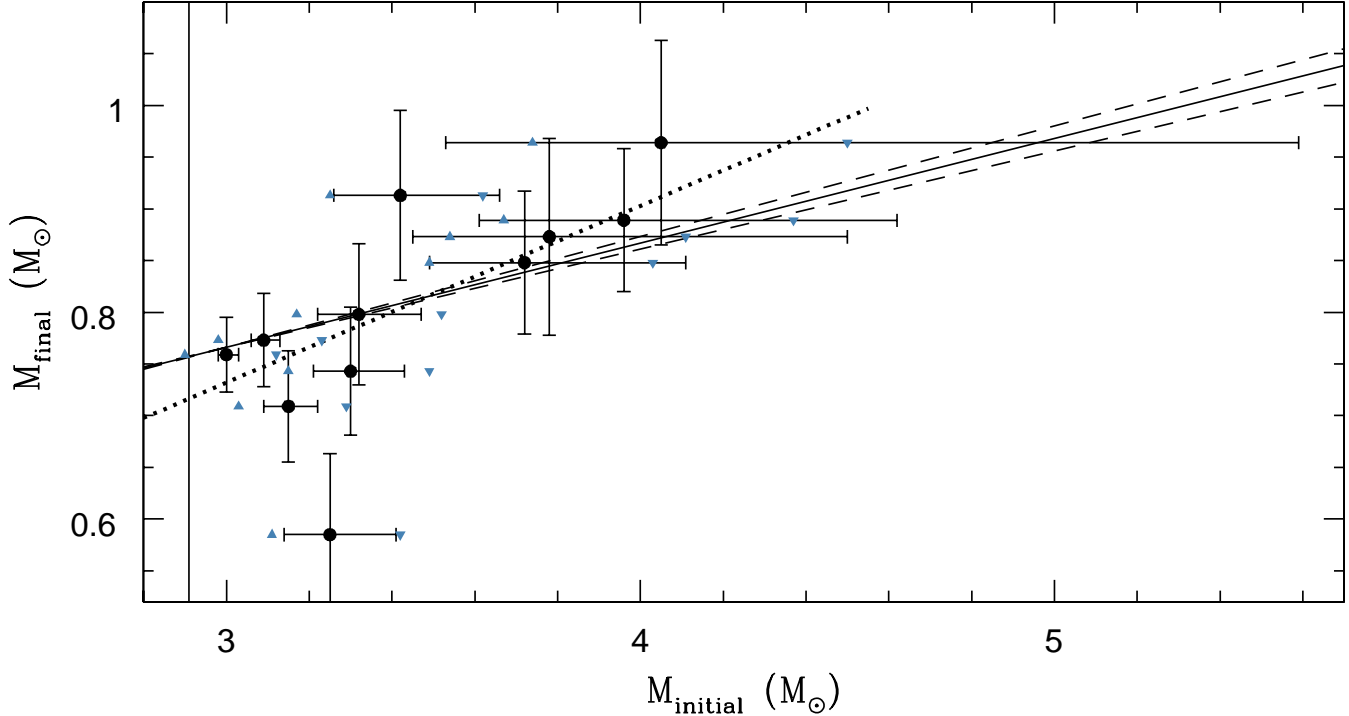}
\end{center}
\vspace{-0.3cm}
\caption{The IFMR relation for our cluster members, adopting a cluster age of 520 Myr. 
The linear IFMR is from K09 (solid line) with its errors (dashed lines).  The steeper 
slope of our linear NGC 2099 IFMR fit (dotted line) is shown.  The solid vertical
line represents the current progenitor mass at the tip of the AGB.  To illustrate 
the effects of adopted age on our final sample, the progenitor masses based on a 470 Myr 
cluster age are shown by inverted gray triangles and the progenitor masses based on a 570 Myr
cluster age are shown by gray triangles.}
\end{figure*}

To further analyze this mass range, in Figure 6 we compare to the large sample of 18 bright and well studied 
white dwarfs from the metal-rich Hyades and Praesepe ([Fe/H]=+0.11$\pm$0.01($\pm$0.03) and +0.16$\pm$0.05($\pm$0.03), 
respectively; Carrera \& Pancino 2011).  The Hyades and Praesepe single white dwarfs have been analyzed using 
consistent techniques and models and are taken from the sample of Kalirai et~al.\ 2014.  
Similar to our NGC 2099 data we fit the initial and final masses of these 18 Hyades and Praesepe white dwarfs using a 
linear relationship that well characterizes the data across this mass range.  We also overplot our final 
sample of white dwarfs from NGC 2099.  In comparison, the Hyades and Praesepe white dwarfs span a comparable 
mass range to NGC 2099, 
but the two older clusters extend to even lower masses.  The linear relations are remarkably 
consistent with nearly identical slopes and no meaningful offset.  This consistency suggests that in this 
mass range and across the metallicity range of these 
three clusters (0$\lesssim$[Fe/H]$\lesssim$0.15) there is no significant metallicity dependence in the IFMR.  This 
is also qualitatively consistent with both the metallicity dependent IFMR models by Marigo \& Girardi (2007) 
and Meng et~al.\ (2008), where there are no meaningful differences ($\Delta$M$_{\rm final}$$<$0.01 
M$_\odot$) predicted at these masses across this specific metallicity range.

\begin{figure}[!ht]
\begin{center}
\includegraphics[clip, scale=0.43]{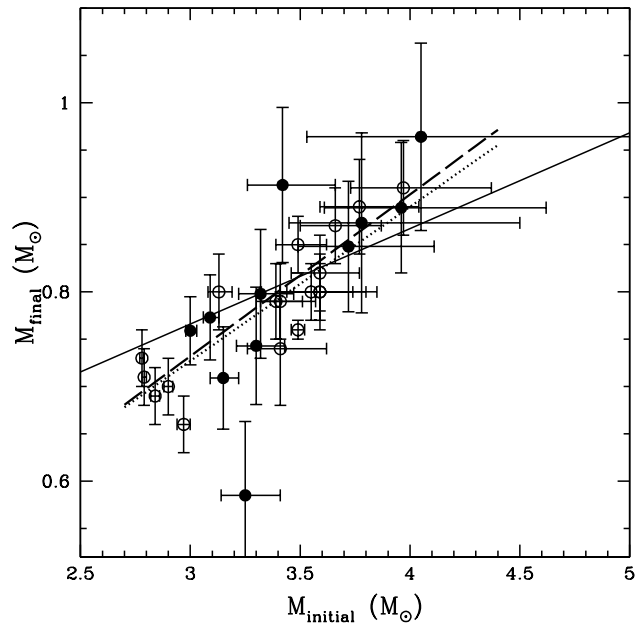}
\end{center}
\vspace{-0.3cm}
\caption{We compare our final sample from NGC 2099 (solid points) and its linear fit (dashed line)
to the white dwarf sample of the Hyades and Praesepe (open circles) with its linear fit (dotted
line).  The slopes are nearly identical and with only an insignificant offset.  As in Figure 5, the shallower 
K09 IFMR is shown in comparison (solid line) but without its errors for additional clarity.}
\end{figure}

The observed consistency between the different samples also can improve our statistics 
by combining the full sample of 29 white dwarfs.  This gives our final combined linear relation 
across the M$_{\rm initial}$ range of 3 to 4 M$_\odot$ of 
M$_{\rm final}$=(0.163$\pm$0.022)M$_{\rm initial}$+0.238$\pm$0.071 M$_\odot$.  Direct 
comparison to the linear relation fit in K09 illustrates that the steeper slope observed 
in this mass range for our data is now far more significant and also meaningfully steeper than
the linear IFMR relations found in Ferrario et~al.\ (2005), Salaris et~al.\ (2009), and Williams et~al.\ 
(2009).  This further indicates that the full mass range of the IFMR cannot simply be 
represented by a linear relation.  

\section{Summary}

We have spectroscopically observed 20 white dwarf candidates in the rich open cluster
NGC 2099.  We have focused on this single open cluster to limit
the potential systematic effects of combining white dwarf data from multiple open clusters, which
may be the driving cause of the large scatter observed in the IFMR.
The rich turnoff and red giant clump of NGC 2099 also allows for a reliable determination of
cluster parameters.  However, to understand the systematics parameter differences that remain 
in the literature, in the Appendix we have considered the color dependence of 
reddening, the effects of metallicity, and the systematic differences between the adopted 
isochrones used to analyze this cluster.

With our spectroscopic parameters determined for the observed white dwarfs, and after 
appropriate cuts based on spectra with large errors, we find that 11 of 14 of the well characterized
white dwarfs are consistent with single star cluster membership.  This is a high percentage relative
to the field and is consistent with the photometric prediction made by K01, and this further emphasizes
the richness of this cluster.  With this rich white dwarf cluster 
population, we have been able to independently test the photometrically determined cluster parameters of distance
modulus and reddening, and both methods give consistent values.  This further 
strengthens our adopted cluster age of 520 Myr.

Applying a 520 Myr age to the 11 cluster white dwarfs provides a well sampled single cluster 
IFMR that spans a white dwarf mass of $\sim$0.7 to 0.95 M$_\odot$, excluding a low-mass outlier,
which corresponds to progenitor masses of $\sim$3 to 4 M$_\odot$.  In general with our white dwarfs, the
observational errors can account well for the observed spread relative to a weighted linear fit
IFMR.  This suggests that stochastic mass loss is not likely a significant contributor to the
spread in the observed IFMR when multiple studies are combined.  The slope for this weighted linear
IFMR spanning initial masses of $\sim$3 to 4 M$_\odot$ is found to be
steep relative to the linear IFMR adopted over a much broader mass range in K09.  

Comparison
of this relation to that found in white dwarfs of the metal-rich Hyades and Praesepe clusters
finds remarkable consistency.  This not only is in agreement with theoretical predictions for
metallicity dependence of the IFMR, but considering together this large sample of 29 white dwarfs 
more reliably establishes the empirical IFMR for progenitors of 3 to 4 M$_\odot$.  
This final relation
provides stronger observational evidence that a nonlinear IFMR, with a shallower slope in higher mass,
may better represent the full mass range.  This plateauing of the relation at higher masses 
further exacerbates the current difficulties in determination of the high-mass IFMR (white dwarf masses 
$\gtrsim$1 M$_\odot$), where there are currently only a limited number of white dwarfs of which all have
large errors.  Therefore, new and higher-mass white dwarfs must be discovered because
extrapolation beyond the current data poorly estimates both mass-loss rates and final
masses for progenitor stars at $>$4 M$_\odot$.

\pagebreak
This project was supported by the National Science Foundation (NSF) through grant AST-1211719.
P-E.T. was supported during this project by the Alexander
von Humboldt Foundation and by NASA through Hubble
Fellowship grant HF-51329.01, awarded by the Space Telescope
Science Institute, which is operated by the Association of
Universities for Research in Astronomy, Incorporated, under
NASA contract NAS5-26555.  Lastly, we thank the referee for the thorough and very helpful report.

\pagebreak
\begin{appendix}
\section{Cluster Parameters}

The cluster parameters of NGC 2099 (M37) play an important role in our analysis, and the
cluster age, distance modulus, and metallicity need to be considered.  In Table 3 we list
the published literature values for the cluster parameters from the past 20 years.  The
determination of all three of these parameters are typically interrelated and also have a strong dependence
on the adopted cluster reddening, which we discuss first.

\subsection{Reddening}
The reddening for this cluster
is moderate, and its past measurements have typical found two ranges: E(B-V) = 0.21 to 0.23 and a 
higher set of values of 0.27 to 0.30.  These reddenings are based on fits of theoretical isochrones 
or of the empirical Hyades main sequence to NGC 2099 photometry, which are typically of only one or
two colors.  In three cases the photometry does not reach deep enough to reliably define the main sequence 
below the turnoff (Mermilliod et~al.\ 1996, Twarog et~al.\ 1997, and Kiss et~al.\ 2001).  While 
the isochrones adopting higher reddening do fit the turnoff of NGC 2099 well, their fit 
of the rich population of red clump giants is poor.  The one exception being Nilakshi \& Sagar (2002; 
E(B-V)=0.30), but they had to adopt a very metal-poor isochrone (Z=0.008) to fit both the turnoff 
and red clump well.  Conversely, isochrones that adopt the lower 
reddening values do noticeably better with the giants and still provide a reasonable but typical
too blue of a fit to the turnoff.

This difficulty of fitting an isochrone simultaneously to both the turnoff and red clump has been 
noted in most previous studies for NGC 2099.  This may be indicative of an issue caused by the typically
assumed solar metallicity or the adopted input physics of the isochrones.  For example, comparing the similar Ventura 
et~al.\ (1998) and Bressan et~al.\ (2012) isochrones to the older Padova group isochrones of Girardi et~al.\ 
(2000) and Bertelli et~al.\ (1994) (see Figure 1) we do find that for the respective best fits the color 
difference between 
the turnoff and red clump is $\sim$0.06 greater in B-V for the latter two isochrones.  A majority of 
this increase is caused by their isochrones predicting a bluer turnoff, which may explain why studies that 
adopt these models also adopt higher reddenings because it is needed to match the observed turnoff, but
as stated this gives a poor red clump fit.  

The narrower color separation found for the Ventura et~al.\ (1998) and Bressan et~al.\ (2012) isochrones still 
has difficulty fitting the turnoff and red clump simultaneously, but this may be explained by the moderate 
reddening of NGC 2099.  The observed reddening is increased in bluer stars, and when using the relation from 
Fernie (1963) the reddening for the turnoff is approximately 10\% higher than that found for the red clump.  
At the moderate reddening for NGC 2099 this factor is no longer insignificant and this diminishes the apparent 
color difference between the turnoff and red clump further by 0.02 to 0.03.  Twarog et~al.\ (1997) also 
discussed the importance of this color dependence on the reddening for these different groups of stars from 
NGC 2099.  Correcting for this greatly improves the simultaneous fit of both the turnoff and red clump seen 
in Figure 1 when adopting the K01 reddening of E(B-V)=0.21$\pm$0.03, which we adopt in this paper.  Lastly, 
to lend further credence to the lower reddenings, private communication with A.\ Steinhauer (2013) gives that 
they find an E(B-V) of 0.22.  Their deep multi-color UBVRI photometry of NGC 2099 provides 
a stronger constraint on the cluster reddening, with the reddening sensitive U photometry in particular.  

\subsection{Metallicity}
The cluster metallicity is quite reddening dependent, both photometrically and spectroscopically.  
A majority of past photometric publications have simply assumed solar metallicity for their isochrone 
fits.  While a handful of photometric metallicity estimates exist, there is large variation ([Fe/H]$\sim$--0.2
to 0.09) and great uncertainty.  Spectroscopically measured [Fe/H] are limited for this cluster, but 
Pancino et~al.\ (2010) derived [Fe/H]=+0.01$\pm$0.05($\pm$0.10) from 3 giants when adopting E(B-V)=0.27 for 
their color-T$_{\rm eff}$ relation.  In contrast, A. Steinhauer (private communication 2013) derived 
a spectroscopic [Fe/H] of --0.136$\pm$0.028$\pm$systematics from 20 main sequence stars using their lower 
reddening of E(B-V)=0.22.  Considering the reddening uncertainty and the corresponding systematic abundance 
error, these metallicities are still consistent.  Reassuringly, when adopting this lower reddening for 
redetermining the T$_{\rm eff}$ for the sample from Pancino et~al.\ (2010), a remarkably consistent [Fe/H] 
of --0.115 is found.  Therefore, when adopting a smaller reddening as we have, both studies find that this 
cluster is consistent with slightly subsolar, and when adopting the higher reddening it is approximately 
solar.  In regard to our comparison to the metal-rich Hyades and Praesepe, a significant reddening
of 0.33 would have to be adopted to give NGC 2099 a consistent high metallicity.  While uncertainty 
does remain, variations of our adopted metallicity from [Fe/H]=--0.1 to solar 
do not meaningly affect our results.  For this paper we continue to adopt solar 
metallicity for NGC 2099.

\subsection{Distance Modulus}
Distance modulus also plays an important role in our white dwarf membership analysis.  However,
unlike the other cluster parameters, a relatively strong consistency is found across previous studies.  
Only in the case of Salaris et~al.\ (2009) is it moderately different, where their high distance modulus 
is the result of the adopted metal-rich isochrone in combination with high reddening.  Here, we 
adopt the distance modulus from K01 of (m-M)$_V$=11.55$\pm$0.13, which based on 
the range in the literature is appropriate.

\subsection{Age}
The age of NGC 2099 from isochrone fits is the cluster parameter with the most significant 
variation found in the previous literature.  However, a majority of this dispersion found in ages 
is simply the result of the adopted isochrones, where the adopted isochrone in each study is 
listed in the notes of Table 3.  We can first discuss
isochrones without convective overshooting, which as is discussed in K01 are found to not reliably fit 
the features of the upper main sequence, and these types of models also give significantly younger 
turnoff/giant fits.  Salaris et~al.\ (2009) also found similar issues
with models that did not include overshoot.  From the studies in Table 
3, only Hartman et~al.\ (2008) used isochrones without overshoot (An et~al.\ 2007) and found 
a 485 Myr age, but when they fit the Yi et~al.\ (2001; hereafter Y$^2$) isochrones with overshoot they found 
an age of 580 Myr. 

Before we begin our systematic comparison of isochrones with overshoot, we note that each isochrone 
set has its own adoption of Z for solar metallicity.  The Bertelli et~al.\ (1994) and Ventura et~al.\ 
(1998) isochrones adopted Z=0.020 for solar, Pietrinferni et~al.\ (2004) adopted Z=0.0198, Girardi 
et~al.\ (2000) adopted Z=0.019, and Y$^2$ adopted Z=0.018.  Systematic differences 
introduced by this variation are negligible and for our solar metallicity isochrones we adopt that 
from each paper.  However, the recent Bressan et~al.\ (2012) PARSEC isochrones adopted a significantly 
lower Z=0.0152 for solar metallicity.  Therefore, for our comparisons we have adopted Z=0.019 for 
their models.  We note that this does cause their isochrone G dwarfs to be systematically
redder (see the central panel of Figure 1), but otherwise this provides a more appropriate 
comparison for systematic differences between these isochrones.

The Bertelli et~al.\ (1994) and Girardi et~al.\ (2000) isochrones have been commonly 
used in the literature age measurements.  Both of these sets of isochrones are from the Padova group, 
and direct comparison of these models at a range of cluster parameters similar to that of
NGC 2099 find that they are nearly equivalent.  Similarly, the isochrones of Pietrinferni 
et~al.\ (2004) were used in Salaris et~al.\ (2009), and Salaris et~al.\ found that the cluster ages
derived from these models have no systematic difference to those of Girardi et~al.\ (2000).  
These three are considered our first group of isochrones.
The other commonly adopted isochrones are those of Ventura et~al.\ (1998) and Y$^2$.
While the Y$^2$ models unfortunately do not include the red clump stars, we find that both of these 
models predict nearly equivalent main sequences, turnoffs, and red giant branches at our range of 
parameters.  We have also compared these isochrones to the most up to date (version 1.2S) 
PARSEC isochrones (Bressan et~al.\ 2012) from the Padova group.  Here, some minor differences are noted
in the lowest-mass main sequence stars and the shape of the turnoff, and with respect to 
Ventura et~al.\ (1998) the red clump stars have subtle differences, but we find these differences
do not significantly affect their relative isochrone fits of NGC 2099.  These three isochrones 
find consistent ages of 520 Myr (using those from Ventura et~al.\ 1998 and Y$^2$) and 540 Myr 
(using those from Bressan et~al.\ 2012).  We consider these three as our second group of isochrones.

Direct comparison of these two different isochrone groups find that there are 
significant systematic differences between their evolutionary timescales.  At ages appropriate for 
NGC 2099, the earlier Padova models estimate ages $\sim$80 to 100 Myr younger than those of Ventura et 
al. (1998), Y$^2$, and Bressan et~al.\ (2012).  Because these up to date Bressan et~al.\ (2012) 
models have superseded the older Padova
isochrones, we take this second isochrone group to provide the better representation of
cluster age.  This systematic age difference can explain the majority of why studies using 
the isochrones of either Bertelli et~al.\ (1994), Girardi et~al.\ (2000), or Pietrinferni et~al.\ 
(2004) consistently found significantly younger ($\Delta$$\sim$100 Myr) ages than K01.

The extremely young age (320 Myr) found in Salaris et~al.\ (2009) and the significantly 
older age (650 Myr) found in Kalirai et al (2005) still require further explanation.  These 
ages are the result of the adopted metallicity in these two studies.  The Salaris et~al.\ (2009) 
isochrones adopted relatively metal-rich isochrones at [Fe/H]=0.09$\pm$0.07.  This higher 
metallicity primarily shifts the cluster redward in B-V by $\sim$0.03, and this in combination with their 
large reddening (E(B-V)=0.3) required that they adopt a significant distance modulus to provide
an appropriate fit to the main sequence.  Hence, an isochrone of an additional $\sim$100 Myr 
younger was necessary to match the now relatively quite bright and blue turnoff.  In contrast to 
this, the previous published work in K05 adopted a metal-poor (Z=0.011) isochrone, and this 
change adjusted their original age (520 Myr; K01) upward to (650 Myr).  Lastly, in regards to the 
effects of metallicity, we should comment on why this did not significantly affect the age found 
by Nilakshi \& Sagar et~al.\ (2002).  In their case the adopted high reddening counteracted the 
blue shifting effect of their low metallicity isochrone, giving both an age and distance 
modulus consistent with the solar metallicity studies.  Lastly, as we have discussed, the most 
recent spectroscopic analyses of NGC 2099 find both Z=0.011 and [Fe/H]=0.09 to likely be too 
extreme, and with our adoption of solar metallicity we continue with the original K01 
age of 520 Myr.

\tablefontsize{\scriptsize}
\begin{center}
\begin{deluxetable*}{l c c c c l}
\multicolumn{6}{c}%
{{\bfseries \tablename\ \thetable{} - Cluster Parameters From Previous Studies}} \\
\hline
Study&(m-M)$_V$&E(B-V)&Metallicity&Age (Myr) & Notes\\
\hline
Mermilliod et~al.\ (1996) & 11.50          & 0.29            & Z=0.02                 & 450        & Bertelli et~al.\ (1994) \\
Twarog et~al.\ (1997)     & 11.55          & 0.27            & [Fe/H]=0.09$\pm$0.07   &  -         & DDO Photometric [Fe/H] \\
Kalirai et~al.\ (2001)    & 11.55$\pm$0.13 & 0.21$\pm$0.03   & Z=0.02                 & 520        & Ventura et~al.\ (1998) \\
Kiss et~al.\ (2001)       & 11.48$\pm$0.13 & 0.29$\pm$0.03   & Z=0.02                 & 450        & Bertelli et~al.\ (1994) \\
Nilakshi \& Sagar (2002)  & 11.6$\pm$0.15  & 0.30$\pm$0.04   & Z=0.008                & 400        & Girardi et~al.\ (2000) \\
Sarajedini et~al.\ (2004) & 11.57$\pm$0.16 & 0.27$\pm$0.03   & [Fe/H]=0.09$\pm$0.07   &  -         & E(B-V) and [Fe/H] are from Twarog et~al.\ (1997) \\
Kalirai et~al.\ (2005)    & 11.50          & 0.23$\pm$0.01   & Z=0.011$\pm$0.001      & 650        & Ventura et~al.\ (1998) \\
Kang et~al.\ (2007)       & 11.4           & 0.21            & Z=0.019                & 450        & Girardi et~al.\ (2000) \\
Hartman et~al.\ (2008)    & 11.57$\pm$0.13 & 0.227$\pm$0.038 & [Fe/H]=0.045$\pm$0.044 & 485$\pm$28 & An et~al.\ (2007) \\
Salaris et~al.\ (2009)    & 12.00$\pm$0.12 & 0.3             & [Fe/H]=0.09$\pm$0.15   & 320$\pm$30 & Pietrinferni et~al.\ (2004) \\
                          & 11.40$\pm$0.12 & 0.23            & Z=0.011$\pm$0.001      & 550$\pm$50 & E(B-V) and [Fe/H] are from K05.\\
Pancino et~al.\ (2010)    & 11.53$\pm$0.19 & 0.27$\pm$0.04   & [Fe/H]=0.01$\pm$0.05   & 410$\pm$155 & Are literature averages other than [Fe/H].\\
\hline
\end{deluxetable*}
\end{center}
\end{appendix}

\end{document}